\def\Interact#1#2{{\cal I}_{#1#2}}

\documentstyle[
12pt,pre,aps
]{revtex}

\begin{document}

\draft

\title{Comparison of the energy histograms and conformations\\
 between different models of copolymers on lattice}

\author{
Edward G.~Timoshenko%
\setcounter{footnote}{0}\thanks{Author to 
whom correspondence should be addressed. 
Phone: +353-1-7162821,
Fax: +353-1-7162127.
Web page: http://darkstar.ucd.ie; 
E-mail: Edward.Timoshenko@ucd.ie},
}
\address{
Theory and Computation Group,
Department of Chemistry, University College Dublin,\\
Belfield, Dublin 4, Ireland}

\author{Yuri A. Kuznetsov%
\thanks{
E-mail: Yuri.Kuznetsov@ucd.ie }}
\address{Centre for High Performance Computing Applications,\\ 
University College Dublin, Belfield, Dublin 4, Ireland}

\author{Roman N. Basovsky
\thanks{
E-mail: Roman.Basovsky@ucd.ie }
}
\address{
Theory and Computation Group,
Department of Chemistry, University College Dublin,\\
Belfield, Dublin 4, Ireland}

\date{\today}

\maketitle

\begin{abstract}
Based on large--scale Monte Carlo simulations on lattice the energy
probability distribution functions are investigated for a large set of 
primary sequences in distinct models of copolymers at low temperatures
below transitions to compacted states. 
Amphiphilic copolymers with hydrophobic and hydrophilic
units are found to produce a single or double peak energy distributions
corresponding to mono- or multi--meric micellar conformations.
However, copolymers with short ranged random `charge' interactions in some cases
are found to produce energy distribution functions with a well
pronounced lowest energy state and a gap separating it from the rest of 
the spectrum. These, however have rather peculiar conformations 
corresponding to effectively immiscible domains comprised from monomers 
of likewise species.
Relevance of these observations for 
coarse--grained models for protein folding is discussed.
$\left.\right.$
\end{abstract}

\section{Introduction}\label{sec:intro}

Synthetic copolymers have numerous industrial applications due to their
rich phase diagram and ability to tailor particular physical
properties by a choice of the monomers used and the order in which they
are polymerised. Generally, a synthesis would produce a quasi--random
arrangement of the monomers, which can be characterised by the probability
distribution of the composition variables $\{ \sigma_i \}$, where 
one agreed to assign $\sigma_i=1$ if the monomer $i$ is of type $a$ and
$\sigma_j=-1$ if the monomer $i$ is of type $b$. The overall composition
is then given by the ratio of the total number of $a$ and $b$ species,
and it would typically be fixed and determined by the ratio of
reactants concentrations during polymerisation, provided that further
separation techniques have allowed one to ensure a good degree
of monodispersity in the degree of polymerisation $N$. 
Likewise, by special procedures one can also
prepare well controlled primary sequences $\{ \sigma_i \}$ corresponding
to di--blocks, tri--blocks, and more generally shorter alternating blocks
of $a$ and $b$ units. Some of perhaps most well studied examples of
synthetic water soluble polymers of these types would include PNIPAM,
PEO-PPO systems and their various modifications \cite{Lindman}.

Most important properties of monomers from the point of view of molecular 
modelling are their degree of hydrophobicity and effective charges, and these
along with the primary sequence would determine the conformations of the
copolymers in solution at given thermodynamic conditions.
Both concepts, however, involve rather complex elementary inter--atomic
interactions: degree of hydrophobicity results from a competition
of the van der Waals, steric, hydrogen--bonding and 
solvent entropic effects, while the
effective charges depend on the electronic properties of the polymer,
solvent dielectric permitivity and degree of screening by the counter--ions
present in solution.

While full--scale atomistic simulations are rather standard for concrete
polymers in solutions nowadays, their ability to study thermodynamic universal
properties of whole classes of macromolecules are rather limited due to 
unbearable computational expenses involved.
For these purposes a number of simplified coarse--grained copolymer
models, which phenomenologically capture the effects of
hydrophobicity and charges, have been proposed and studied both
analytically \cite{Sfatos-93,Pande-94,Thiru-96,OurRandom}
and by Monte Carlo simulation (see e.g. \cite{MonCarlo} and
references therein), often on lattice, which further significantly
reduces computational expenses.

A great deal of related research has also been undertaken in relation
to the problem of protein folding in the context of random heteropolymer
models of proteins \cite{Sfatos-93,Pande-94}. 
Although natural proteins are composed of
20 types of amino acid residues with rather refined properties, in a
simplified model one can roughly divide all amino acid residues into
polar (hydrophilic) and apolar (hydrophobic),
as well as into effectively charged and neutral.
This broad division produces 2-letter models of proteins, and since the
primary sequence of each protein is unique, applied to a whole ensemble 
of sequences this leads to random heteropolymer models of proteins.

Depending on which particular feature one is trying to describe thus
there are models of amphiphilic \cite{Garel,Clusters} and charged
copolymers \cite{Sfatos-93} after solvent molecules have been effectively
excluded from the consideration by integration (see Refs. 
\cite{Messina,CopStar} for details).

The first model has a linear term in composition
variables, $(\sigma_i+\sigma_j) V(r_{ij})$, in the Hamiltonian, so that 
hydrophobic units $a$ effectively attract each other (and thus dislike solvent
contacts), hydrophilic $b$ units effectively repel each other 
(and thus like solvent contacts), whereas hydrophobic-hydrophilic $a\,b$
direct pair--wise interactions are zero. Such model will be referred to
as hydrophobic--hydrophilic or HH model. As a simple variation of this
we can also introduce hydrophobic-ideal or HI model, in which $b$ species
have zero repulsion from each other and thus effectively behave as ideal.

The model with a quadratic term in composition
variables, $\sigma_i\,\sigma_j V(r_{ij})$, in the Hamiltonian is 
often called the random `charge' model.
In this context one can introduce the true charge or briefly CA model,
in which likewise charges repel and opposite charges attract each other,
or the model in which the quadratic term appears in the Hamiltonian
with the opposite sign, so that likewise charges attract and opposite
ones repel each other, which is naturally to call the anti--charge
or AC model.  Although the latter model does not describe true
atomistic charges, such terms nevertheless
appear to reproduce complex non--local effective monomer--solvent 
interactions arising after coarse--graining
of models with complex intra--molecular potentials. Namely, the AC model
was most popular in literature in analytical and lattice Monte Carlo
studies of random heteropolymer models of proteins \cite{Sfatos-93,Pande-94}. 
As we shall see later it indeed has the most peculiar properties, and although
its applicability to synthetic copolymers is least justified, no single model
is good enough for describing the complex behaviour of proteins.

In this work we would be interested to extensively investigate the energy
probability distribution functions $P_T(E)$ 
in the HH, HI, CA and AC
models for a large set of different heteropolymers sequences
at low temperatures after transitions to compacted states, whether
crystalline or glassy, as well as to try to relate the observed
distributions to copolymer conformations, by which we understand
the arrangement of monomers in relation to each other in space.

The energy probability distribution is formally defined as,
\begin{equation}\label{eq:PE}
P_T(E)=\frac{1}{Z_T}\sum_{m} \exp\left(-\frac{H_m}{k_B T}\right) \delta(E-H_m),
\qquad Z_T \equiv \sum_{m} \exp\left(-\frac{H_m}{k_B T}\right),
\end{equation}
where $m$ refers to microstates of the system, and this is linearly related
to the density of states times an explicit exponential factor and a
constant $Z_T$,
\begin{equation}\label{eq:NE}
{\cal N}(E)\equiv \sum_{m} \delta(E-H_m) = P_T(E)\,
\exp\left(\frac{E}{k_B T}\right)\,Z_T.
\end{equation}

We note that although in Ref. \cite{Socci} Monte Carlo histogram technique
was successfully applied to a 27-mer on a cubic lattice
without enumerating all compact 3x3x3 cube states, here we shall 
be able to investigate many copolymer sequences of a longer chain 
on a large lattice with a 
good statistical sampling able to distinguish universal patterns
of behaviour in the four aforementioned copolymer models directly  
at low temperatures without using extrapolation by means of Eq. (\ref{eq:NE}).
This was previously problematic as a much larger sampling is required
at temperatures below the transitions to compacted states.

\section{Model}

We adopt the Metropolis technique in the lattice model of our
Ref. \cite{MonCarlo}.
This model, apart from the connectivity and excluded volume constraints,
includes short--ranged pair--wise interactions between
lattice sites.

In case of copolymers one has to distinguish two types
of monomers {\it a} and {\it b} and ascribe to them
three values of Flory interaction parameters as follows \cite{MonCarlo},
\begin{eqnarray}
\chi_{aa} & = & \frac{2 \Interact{s}{a} - \Interact{a}{a} -
                \Interact{s}{s}}{k_{B}T},
\quad
\chi_{bb}  =  \frac{2 \Interact{s}{b} - \Interact{b}{b} -
                \Interact{s}{s}}{k_{B}T}, \label{chibb} \quad 
\chi_{ab}  =  \frac{  \Interact{s}{a} + \Interact{s}{b} -
                \Interact{a}{b} - \Interact{s}{s}}{k_{B}T}. \label{chiab}
\end{eqnarray}
As in Ref. \cite{CopStar} we shall use the parametrisation 
such that $\sigma_a=1$, $\sigma_b=-1$ and
\begin{equation}
\chi_{ij} = \chi_0 + \Delta\frac{\sigma_i+\sigma_j}{2}
\ 
\mbox{(HH, HI)}, \quad
\chi_{ij} = \chi_0 + \Delta \sigma_i \, \sigma_j
\
\mbox{(CA, AC).}
\end{equation}

Thus, all interaction parameters can be summarised as in Table \ref{tab}.
We may note that the hydrophobic--hydrophilic and hydrophobic--ideal models
use the same formal expression, but different values of $\chi_0$ and 
$\Delta$.

\section{Results}
\label{sec:results}

In our study we have considered copolymers of length $N=48$ units
with equal number of $a$ and $b$ monomers in each sequence on a large enough
cubic lattice of linear dimension $L=50$ in order to make the influence
of boundary conditions negligible. 
For each of four models
we have analysed a number of periodic and 1000 randomly generated sequences.
Every initial random walk conformation was first subjected to a
slow quasistatic equilibration corresponding to gradually decreasing
temperatures until the values of $\chi_{ij}$ presented in Table \ref{tab}
have been reached. This procedure is similar to `simulated annealing'
and permits to overcome excessive trapping of the system in metastable
states. Moreover, to be able to deal with the issue of 
`non--ergodicity' this cooling was applied to a fairly large ensemble 
(typically over 200) of different initial conditions.
To obtain probability distributions of various observables, such as e.g.
energy, we then performed hundreds of measurements separated by large
number of Monte Carlo sweeps (i.e. typically hundred of thousands of attempted
Monte Carlo moves), resulting in the 
total number of independent  statistical measurements equal to 
20000 in each case. Finally, normalised energy histograms provide Monte Carlo
approximations to the exact energy probability distribution functions $P(E)$
defined by Eq. (\ref{eq:PE}).

Due to the enormous amount of data obtained and obvious constraints of 
the paper size we shall only discuss here most interesting and representative
results, while the whole data set is available online \cite{Web} for
an interested reader.

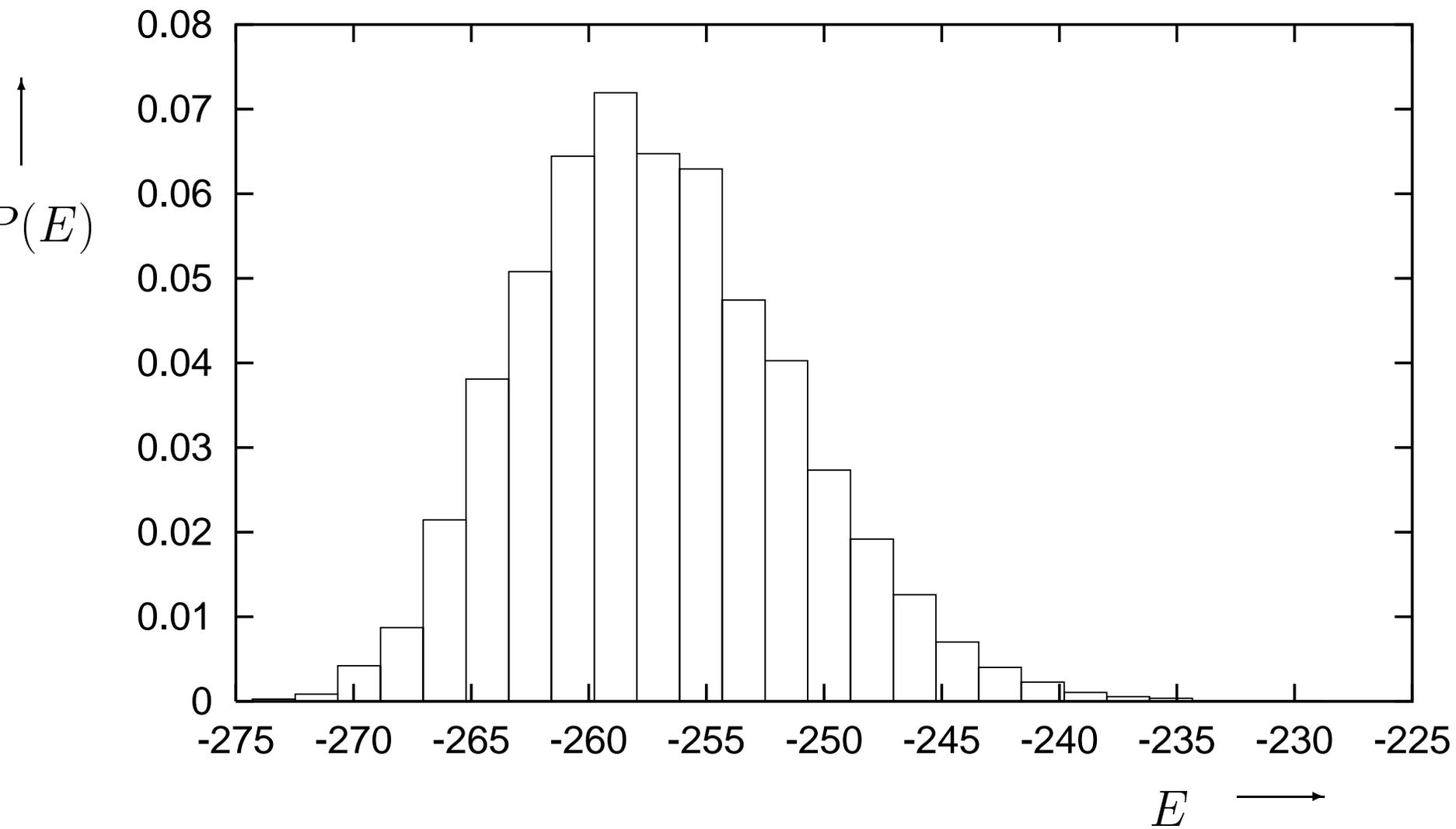
\begin{figure}
\caption{
Energy distribution function for the random sequence (s1)
{\it abbbbabaabbaababababbbaabababbbabbbbaaaababaaaaa}
in the hydrophobic--ideal model. Here and below $k_B T$
units of energy are used.
}\label{fig:1}
\end{figure}

A typical energy distribution function of HI model shown in Fig. \ref{fig:1} 
has a bell shape. Note that low energy states contribute little to this
function because the density of states ${\cal N}(E)$  is rapidly
decreasing with decreasing energy $E$.
The general outlook of $P(E)$ changes
little for different sequences, with only mean value $\langle E\rangle$
and width $\langle \left(E-\langle E\rangle\right)^2\rangle^{1/2}$
of the distribution being sensitive on the sequence.

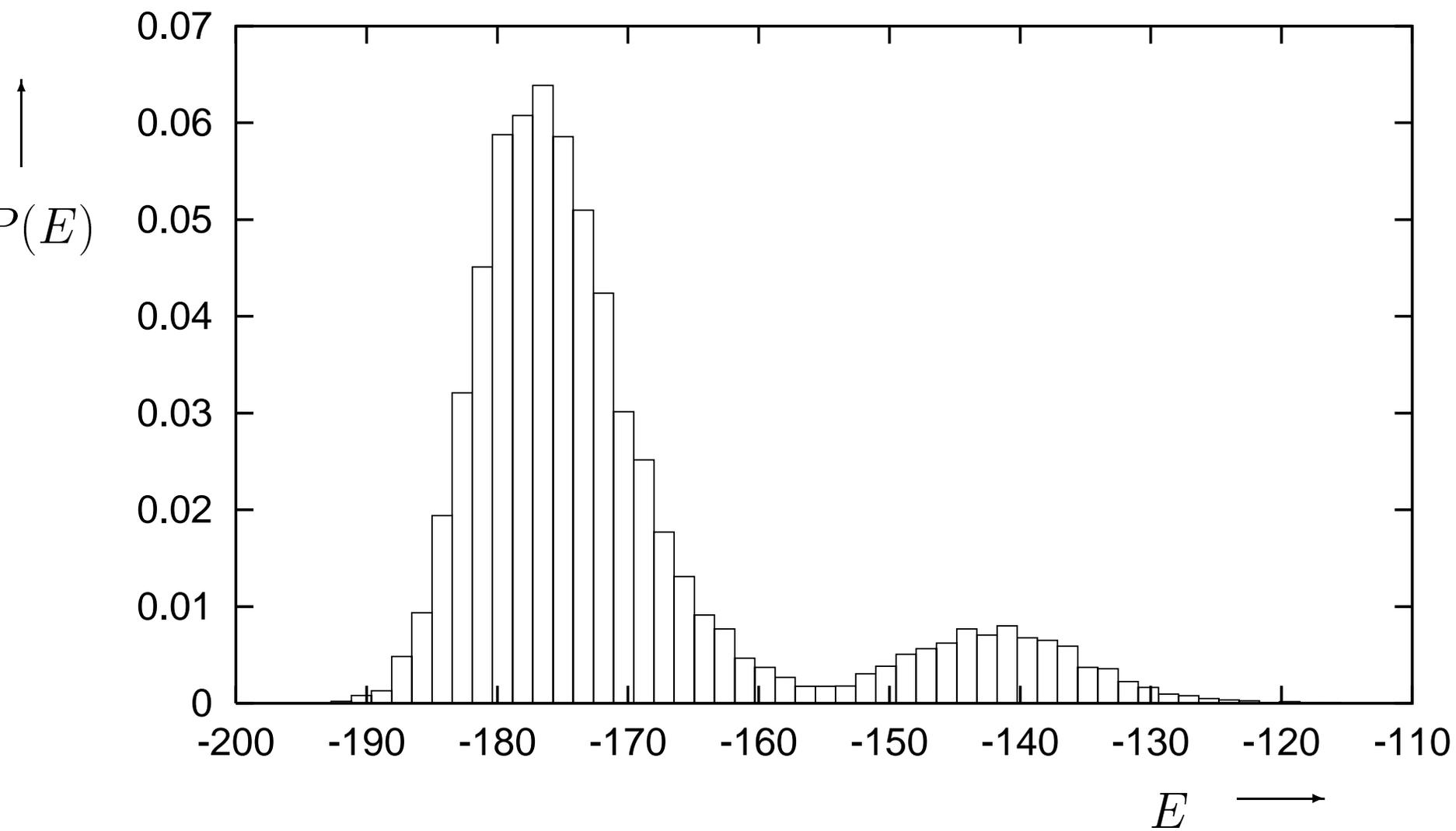
\begin{figure}
\caption{
Energy distribution function for the random sequence (s2)
{\it ababbaabaababaabababbbabbabaabababaabbababbbaaba}
in the hydrophobic--hydrophilic model.
}\label{fig:2}
\end{figure}

For most sequences the energy distribution function in the HH model 
is similar to that of the HI model. However, a few percents of sequences
in the HH model also possess a second peak in $P(E)$ as shown in 
Fig. \ref{fig:2}. The secondary peak at higher energies is considerably
lower and could be simply understood by visualising typical copolymer
conformations corresponding to a given energy distribution.
First of all, a snapshot in Fig.~\ref{fig:3}a shows a typical
3-D polymer shape for a single peak situation. A clear feature of
both the HI and HH models is the micro--phase separation (MPS) of the
globule onto a hydrophobic (black) core and a hydrophilic (gray)
shell as in a micelle. As for the case of a sequence with a two--peak
$P(E)$ function, this would correspond to a dimer (or a multi--mer for longer
chains) of sub-globules connected by a predominantly hydrophilic bridge.
Obviously, such a situation can be realised only for sequences which
do have a fragment in their primary sequence with a predominantly 
hydrophilic units.

The energy distributions in the CA model also have a trivial single
peak shape as in Fig. \ref{fig:1}. However, the polymer conformations are
rather different in this case. Thus, in Fig.~\ref{fig:4} one can see that
although the globules are still compact, the internal arrangement
of the monomers within are rather peculiar and have an alternating 
rather than a core--and--shell structure. This arrangement obviously
tends to maximise the number of $a\ b$ contacts favourable in this model.
Figs.~\ref{fig:4}a and b corresponding to a random and a long blocks sequences
respectively show, however, that there is no tendency of forming better structured
conformations (in the sense of maximising the MPS
between $a$ and $b$ units) for periodic sequences with increasing
block length as was the case in the HH and HI models. 

Most interesting energy distributions of the four considered models occur
in the AC model. Although most sequences here would still have a single
shape distribution as in the other models, there are a few percents of
sequences in the AC model which have energy histogram as shown in
Fig.~\ref{fig:5}. These sequences possess a highly populated 
visible lowest energy state, which is often separated by a gap
from the rest of the spectrum. This may be rationalised as occurring due to
much a higher density of states ${\cal N}(E)$ at low energies. 

This unusual behaviour of $P(E)$ also corresponds to interesting
conformations of the copolymer. In Figs.~\ref{fig:6}a and \ref{fig:6}b
these are shown for examples of periodic sequences of different block
lengths. For long blocks sequence in Fig.~\ref{fig:4}a there is a clear
dumbbell shape with $a$ and $b$ units in their separate homopolymer--like
subglobules avoiding unfavourable contacts with each other.
For shorter blocks sequences as in Fig.~\ref{fig:4}b this separation of
$a$ and $b$ units can not be as perfectly realised due to the chain 
connectivity constraints, thus producing merely a multi--domain globule of
homopolymer--like clusters. Also, the increase of the block length
shifts the $P(E)$ function towards lower energies due to a better 
optimisation of competing pair--wise interactions.


\section{Discussion}

As we have seen here the energy probability distributions are
bell shaped and shifted towards low energies in most cases.

The second peak in the HH model occurring for some sequences, 
especially those having a  long hydrophilic segments in the chain,
corresponds to formation of di--mers and multi--mers.
Naturally, the HI model has similar features to HH model, although it has
a lower tendency for forming multi-mers.
Speaking of the conformations,
only the HH and HI models produce good micro--phase separation (MPS) with a 
predominantly hydrophobic core and hydrophilic surface.
The degree of MPS increases with increasing a characteristic repeating
block in a sequence. 

Conformations in CA model do not have a MPS structure
and are rather complex amorphous (glassy) globules.
The characteristic block length has little effect on the shape
and energy in this model.

The AC model has most interesting and most sequence--dependent
energy distributions of all. It has a well populated lowest
energy state separated by a gap from higher energy states for certain
sequences, which perhaps can be viewed as {\it good folding sequences}
according to Ref. \cite{Socci}.
This shape is generally believed to provide good accessibility
and stability of the lowest energy state
identified as an analog of the native state in proteins.
We may note that this result agrees with Fig. 2 in Ref. \cite{Socci},
in which the AC model was in fact studied at somewhat different values of
parameters as compared to Table \ref{tab}.

However, the AC model exhibits phase separation of $a$ and $b$ species
similar to two immiscible liquids under connectivity constraints,
which is rather different from $a$-surface and $b$-core
MPS in HH model. Increasing typical block length here leads to 
dumbbell structures rather than micelles, which also has an effect of
lowering the energy considerably. Short block sequences produce
multi--domain globules of identical species.

Thus, while AC model has an energy histogram which bears
resemblance to that of a protein in the native state, it has
conformations which have little to do with typical protein shapes,
especially in terms of MPS.
The HH model, on the other hand, has conformations similar
to those of proteins, but rather trivial energy histogram with
easily accessible low energy states and no barrier or special
pathway to these states from the coil.
Thus, unfortunately none of the commonly used coarse-grained
lattice models bears any direct relevance for modelling proteins,
at least at the level of 2-letter models.

\section{Conclusion}

We have presented results for four distinct models of
copolymers on lattice obtained from Monte Carlo simulations
of a large ensemble of 20000 independent initial conditions
for 1000 randomly generated and a number of regular
sequences with equal ratio of
$a$ and $b$ monomers. Features of energy distribution functions (histograms) 
and corresponding polymer shapes (conformations) are discussed
in the hydrophobic-hydrophilic (HH), hydrophobic-ideal (HI),
charge (CA) and anti-charge (AC) models of copolymers at temperatures
much below transitions to compacted states.
Most sequences in all of models have a single bell shaped energy
probability distributions.

However, some sequences in the HH model have two peaks distributions.
Moreover, a number of sequences in AC model have more than two
peaks and often a gap between the lowest energy and higher energy states.
Conformations and micro--phase separation of $a$ and $b$ species corresponding
to distinct distributions in these models are also analysed and these results
may be of interest for a range of synthetic polymers in solution.

Finally, from our analysis it appears that while separate features expected
from proteins can be represented by the HH and AC models of copolymers, neither
model possesses all of these features together.
Thus, a proper continuous space treatment with the full account of Coulomb,
van der Waals, bonded and other specific interactions may
be necessary to capture even in a minimal way the folding behaviour
of proteins.

\acknowledgments

The authors acknowledge most interesting discussions with 
Professors F.~Ganazzoli, T.~Garel, H.~Orland, and P.G. Wolynes
at different stages of the current research.
This work was supported by grants SC/99/186
and BC/2001/034 from Enterprise Ireland.





\newpage
\section*{Figure Captions}

\begin{figure}
\caption{
Snapshots of typical copolymer conformations for sequence (s2) in the 
hydrophobic--hydrophilic model. Figs. (a) and (b) correspond to
$E=-172.2$ (left peak) and $E=-135.8$ (right peak)  
in Fig. \ref{fig:2} respectively.
}\label{fig:3}
\end{figure}

\begin{figure}
\caption{
Snapshots of typical low energy copolymer conformations in the 
charge model. Figs. (a) and (b) correspond to
random sequence (s3) {\it aababbbbabaababaaabbababbaabaabaaaaabbbbbabbabab}
and periodic sequence $(a_{12}b_{12})_2$
respectively.
}\label{fig:4}
\end{figure}

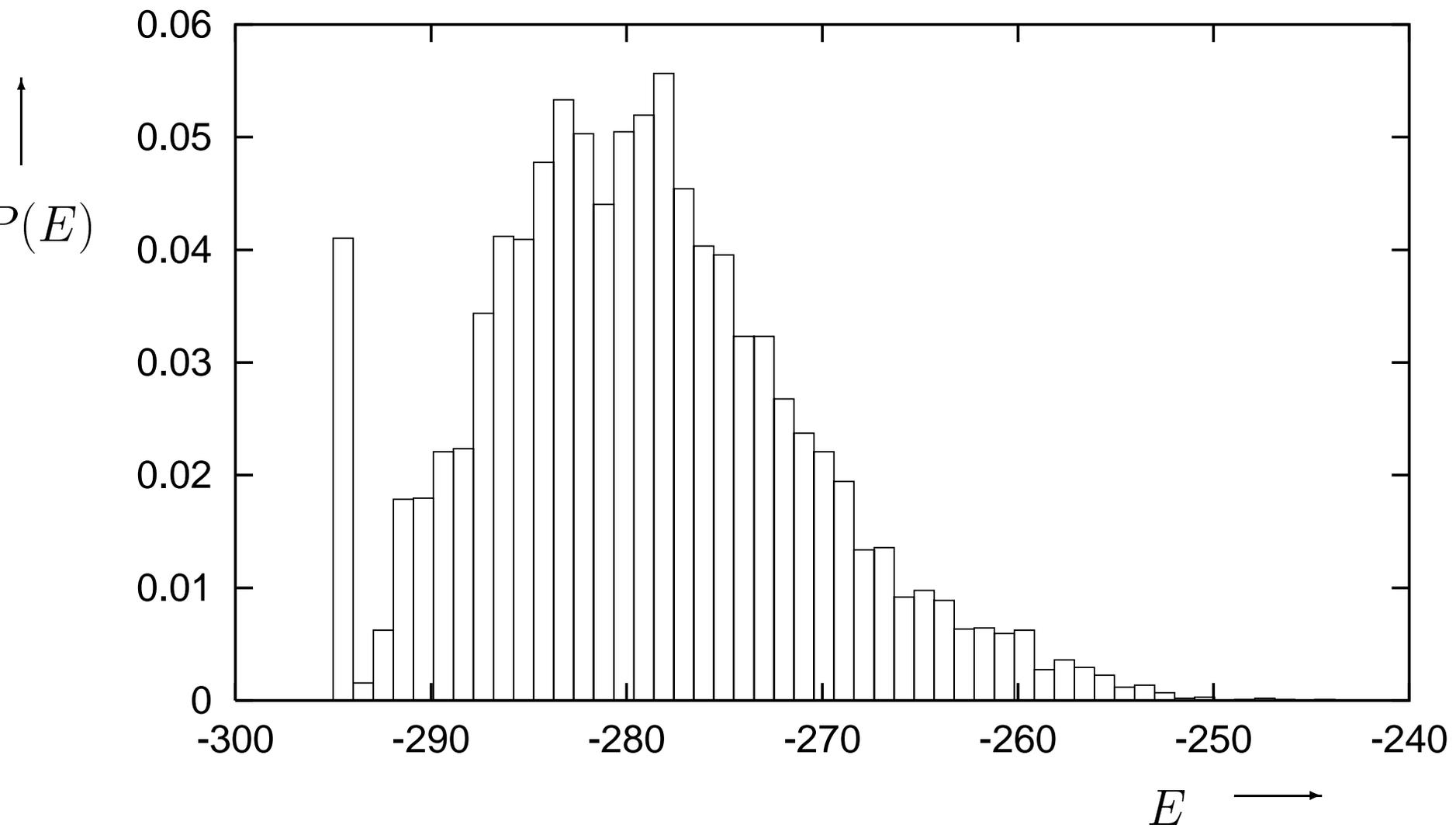
\begin{figure}
\caption{
Energy distribution function for the random sequence (s4)
{\it ababbabaababbaaaaababbabbaaaabbabbaabbbaabbbbaab}
in the anti--charge model.
}\label{fig:5}
\end{figure}

\begin{figure}
\caption{
Snapshots of typical low energy copolymer conformations in the 
anti--charge model. Figs. (a) and (b) correspond to
sequences $(a_{12}b_{12})_2$ (at $E=-376$) and $(a_{3}b_{3})_8$ 
(at $E=-324$) respectively.
}\label{fig:6}
\end{figure}

\newpage

\section*{Tables}

\begin{table}
\caption{Values of the Flory interaction constants
for four different models of copolymers on lattice.
}\label{tab}
\begin{tabular}{lccccc}
Model & $\chi_{aa}$ & $\chi_{ab}$ & $\chi_{bb}$ & $\chi_0$ & $\Delta$ \\\hline\hline
HI    &     2       &     1     &     0     &    1   &    1   \\
HH    &     2       &    0.4    &   -1.2    &   0.4  &   1.6  \\
CA    &     0       &     2     &     0     &    1   &   -1   \\
AC    &     2       &     0     &     2     &    1   &    1   \\
\end{tabular}
\end{table}

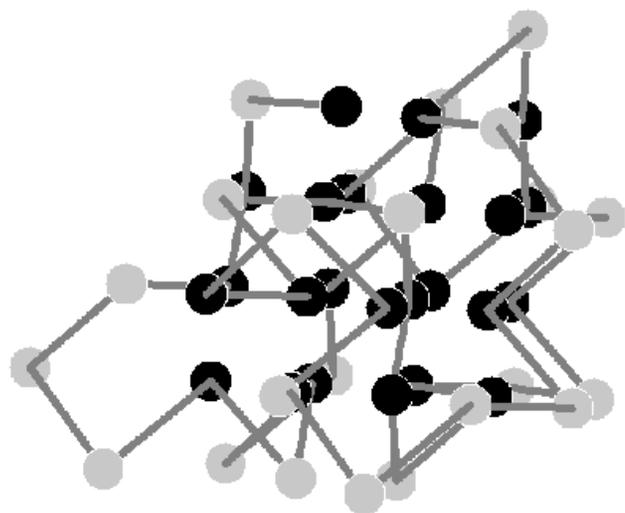
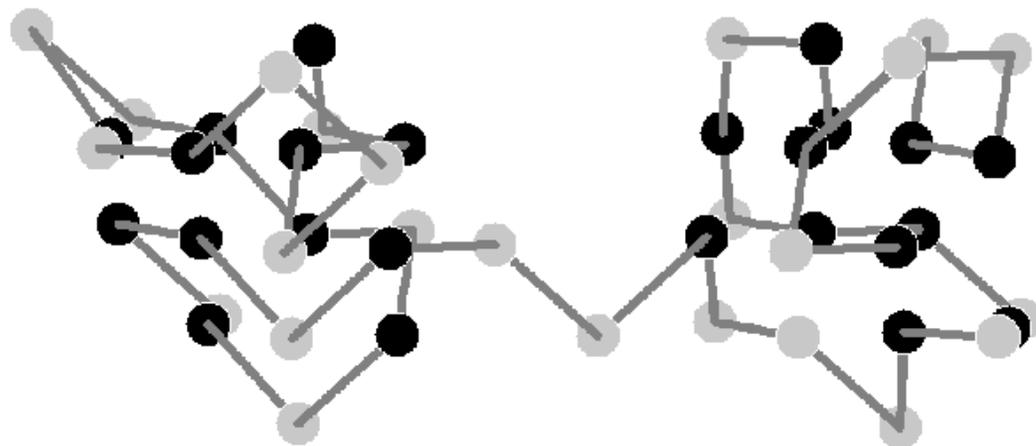
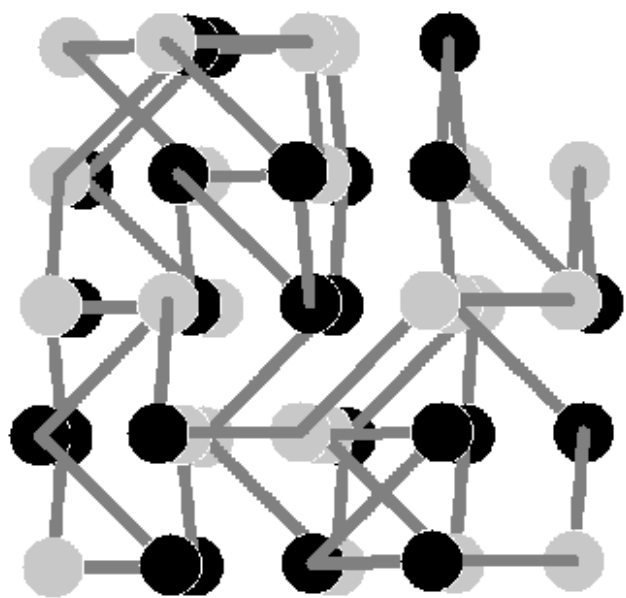
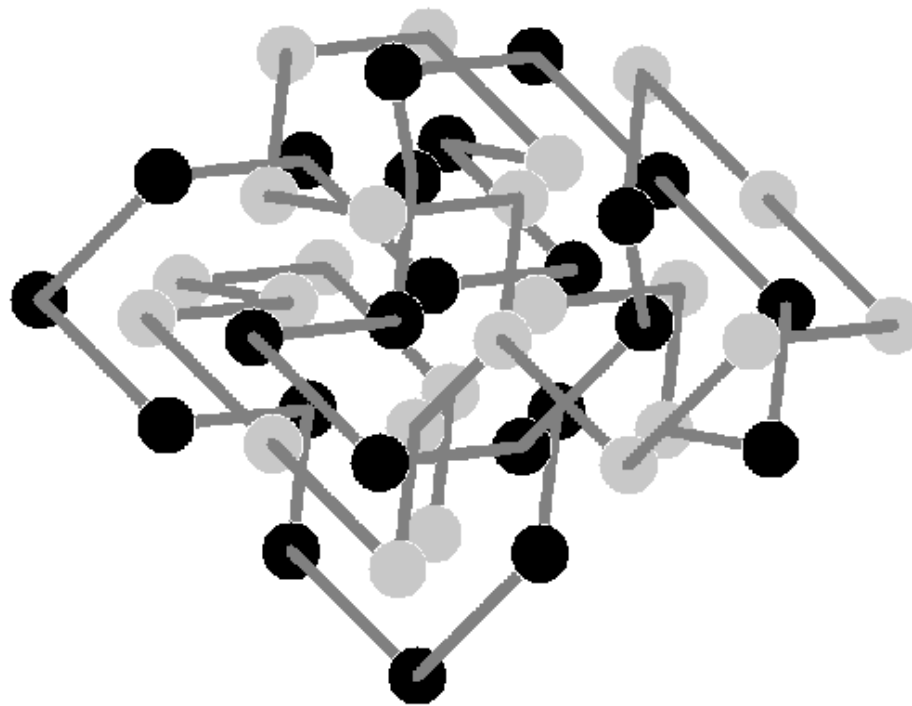
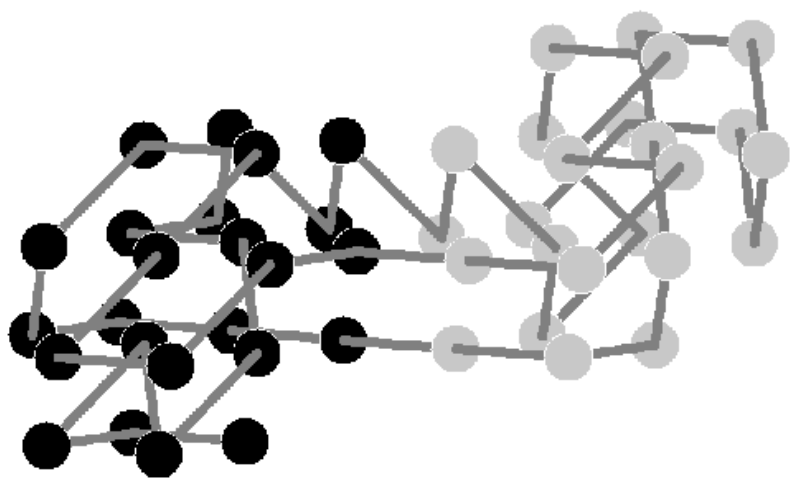
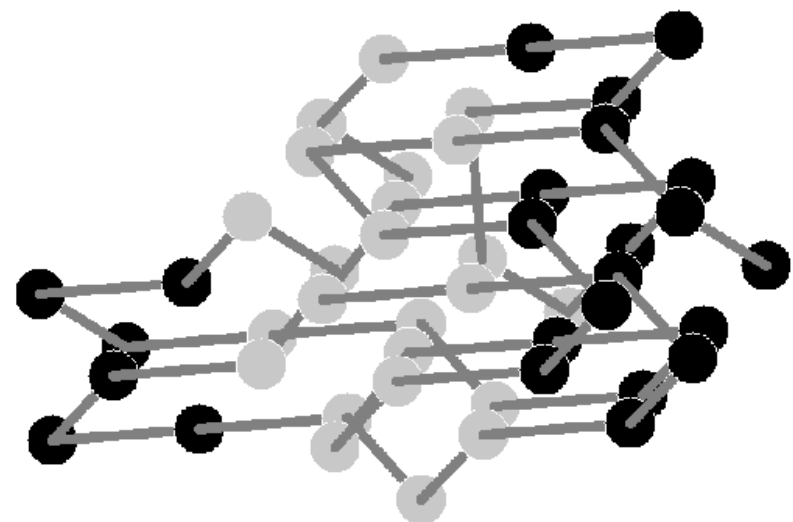
\end{document}